# GENESIS OF A PYTHAGOREAN UNIVERSE

Alexey Burov* and Lev Burov°

*FNAL, Batavia, IL

°Scientific Humanities LLC, San Francisco, CA

*Abstract*

The full-blown multiverse hypothesis, chaosogenesis, is refuted on the grounds of the large scale and high precision of the already discovered laws of nature. A selection principle is required not only to explain the possibility of life and consciousness, but also theoretizability of our universe. The weak anthropic principle provides the former, but not the latter. As chaosogenesis is shown to be the only thinkable scientific answer to the question of why the laws of nature are the way they are, its refutation means that this question cannot be answered scientifically.

## Introduction

The task of science, as it is generally assumed, is to find the laws of nature allowing both to explain the diversity of observations as well as to predict new ones. Science seeks to discover the logic that is hidden beneath phenomena and which determines their flow and qualities. The understanding of truth as uncovering of hidden essence, as dis-covery, is embedded in the Greek word αλήθεια (truth), consisting of negation (α-) and λήθη, which means a veil or concealment. Pythagoras taught that this essence is the harmony of hidden unity which can be expressed in the language of numbers. When Galileo stated that nature is a book written in the language of mathematics, he was expressing this ancient Pythagorean credo. The same can be said about Dirac, whose fundamental belief was that "the laws of nature should be expressed in beautiful equations", and about Einstein who believed that the strongest and noblest motive for the scientific search is a deep conviction of the rationality of the universe, saturated with the cosmic religious feeling.

When theories that exhaust phenomena are formulated and logically unified into a single theory of everything, the task of fundamental science is finished. Whatever this theory of everything may be, other theories in physics will be its consequences as limit cases or asymptotes. Although humanity does not now and may possibly never have such a theory in its fullness, many of its limit cases are known to us as concrete theories, such as classical and quantum mechanics, general theory of relativity, the standard model, and others.

The laws of nature are discovered as composite and specific mathematical structures. As these structures are revealed, we unavoidably come to a certain question regarding the structures themselves. First of all, why does any law expressed by one or another mathematical formula structure our world at all? While it is thinkable for a universe to be structured by any logically consistent system, out of this infinite set of structures only one determines our universe. Why this structure and not another? Why are the laws simple enough to be discovered? Why are they mathematically beautiful? Who or what singled it out and on what ground?

In this way the laws of nature become a problem, though not in the usual scientific context of searching them out, but as something that requires its own explanation. The illusory nature of an explanation that does not go beyond natural laws was pointed out by Ludwig Wittgenstein [1]:

> The whole modern conception of the world is founded on the illusion that the so-called laws of nature are the explanations of natural phenomena. Thus people today stop at the laws of nature, treating them as something inviolable, just as God and Fate were treated in past ages. And in fact both are right and both wrong: though the view of the ancients is clearer in so far as they have a clear and acknowledged terminus, while the modern system tries to make it look as if everything were explained.

Here Wittgenstein criticizes a silent acceptance of a composite and special mathematical structure as the ultimate explanation of the world. Such explanation barred from further questioning and not subject to reasonable ground of its own existence is an affirmation of unreasonableness of this ground. In other words, it is an acceptance of absurdity as the ultimate foundation of existence, or, in the words of Paul Davies [2]:

> One can ask: Why that unified theory rather than some other?... Why a unified theory that permits sentient beings who can observe the moon? One answer you may be given is that there is no reason: the unified theory must simply be treated as "the right one," and its consistency with the existence of a moon, or of living observers, is dismissed as an inconsequential fluke. If that is so, then the unified theory— the very basis for all physical reality— itself exists for no reason at all. Anything that exists reasonlessly is by definition absurd. So we are asked to accept that the mighty edifice of scientific rationality — indeed, the very mathematical order of the universe— is ultimately rooted in absurdity!

Such superstition destroys the meaning of fundamental science by undermining the importance of reason, subjected by this superstition to the absurd.

What can be the answers concerning the source of the laws of nature? Is there any way of choosing or rejecting one or another? That is the topic of discussion in the present article.

## The Fine Tuning Question

"There is now broad agreement among physicists and cosmologists", writes Paul Davis [3], "that the universe is in several respects 'fine-tuned' for life". Similarly, Stephen Hawking has noted:

> The laws of science, as we know them at present, contain many fundamental numbers, like the size of the electric charge of the electron [*fine structure constant*] and the ratio of the masses of the proton and the electron. ... The remarkable fact is that the values of these numbers seem to have been very finely adjusted to make possible the development of life [4].

Another crucial point is articulated by Alexei Tsvelik [5]:

> [since] the number of existing life-imposing conditions by far exceeds the number of constants, their fulfillment could not be achieved by fine tuning of these constants and required also the right choice of the fundamental principles of physical laws.

The premise of the fine-tuned universe revived the old metaphysical problem of the source of order in the world as the problem of fine-tuning: who or what tuned the universe so fine? A pure scientific approach required finding an objective answer: not "somebody" but "something" as the cause of tuning.

## Order From Chaos

It is thought that this "something" could be any combination of laws of nature provided by one or another general theory and random factors; or, using the terms of Platonic philosophy, any combination of forms and chaos. However, as it was noted by Wittgenstein, any theory used in that respect itself requires to be explained. John A. Wheeler expressed the same as a question: why is this very theory structuring everything existent? Why doesn't some other theory instead? In other words, the use of any theory for this does not solve the problem of fine-tuning, but moves it to a higher level. The only way to solve this problem totally in the framework of science is to show a possibility of appearance of being from nothing, or *chaosogenesis*, the appearance of order from chaos. Indeed, theories, being specific formal structures, are limited and composite entities, and thus lead to the question "why this theory and not other?". Chaos per se is limitless and structureless, a totality intrinsically undivided into "this" and "that", whose various manifestations differ from each other due only to the variety of doors that one or another theory opens for chaos to enter.  Historically, the idea of chaosgenesis is very old, having been traced down to Hesiod and pre-Socratics, and it had been opposed by the Pythagoreans and Platonics. For instance, Plotinus wrote: "Any attempt to derive order, reason, or the directing soul from the unordered motion of atoms or elements is absurd and impossible."[6] Not all contemporary cosmologists share Plotinus' views on the chaosgenesis, so the idea is frequently pronounced.

Max Tegmark has formulated the "Ultimate ensemble theory of everything", whose main motivation is clearly expressed [7]:

> If the TOE [*theory of everything*] exists and is one day discovered, then an embarrassing question remains, as emphasized by John Archibald Wheeler: Why these particular equations, not others? Could there really be a fundamental, unexplained ontological asymmetry built into the very heart of reality, splitting mathematical structures into two classes, those with and without physical existence? After all, a mathematical structure is not "created" and doesn't exist "somewhere". It just exists. As a way out of this philosophical conundrum, I have suggested that complete mathematical democracy holds: that mathematical existence and physical existence are equivalent, so that all mathematical structures have the same ontological status.

Thus, for Tegmark the terminus ultimately explaining everything existing is the totality of all mathematical forms, the platonic world. To "just exist", the mathematical structure has to be self-consistent, logically acceptable. What he doesn't mention is the unity of these forms. This unity must not only somehow bind every one of them together but it has to  guarantee their self-consistency. The forms though are mental entities. They are not thinkable without a mind which contains them as truly self–consistent. Thus, we have to conclude that this unity, the terminus of Tegmark's being, is an absolute mind, even if it is not mentioned at all. What makes this mind special and distinctive from its various platonic versions is its total indifference to the forms it contains. That is what Tegmark calls "the mathematical democracy".

It has to be stressed, that purely by itself, without any forms involved, chaos cannot produce anything, and Tegmark's model is not an exception from this rule: it assumes that all possible worlds are based on mathematical structures, such as groups, algebras, fields, sets of equations, and other formal systems. It also assumes that there is a way for these structures to show themselves as phenomena, and to be observed both as mathematical and physical objects. Chaos comes in this picture as a randomness of a universe we happen to be born in, with  the only limitation that the laws of this universe are compatible with life and consciousness.  What makes Tegmark's model very special is its minimal involvement of a priori concretization or selection principles, which is why we are equating this model of "mathematical democracy" with  chaosogenesis.

A possibility for the structure of the fundamental laws of nature to be random to some unclear degree and beyond that to be non-randomly selected by some unpronounced entity was expressed by several leading scientists, e. g. by Andrei Linde (see a citation below) and Steven Weinberg [8]:

> …we have to keep in mind the possibility that what we now call the laws of nature and the constants of nature are accidental features of the big bang in which we happen to find ourselves, though constrained (as is the distance of the Earth from the Sun) by the requirement that they have to be in a range that allows the appearance of beings that can ask why they are what they are.

The Darwinian theory of evolution is widely believed to explain the birth of order from chaos. To follow its line of thought, our universe is considered a member of a huge or infinite ensemble of universes, one generated by the other, with daughter universes mostly inheriting the logical structure of the mother ones, adding some mutations on top [9,10]. After the heredity and variation of the multiplying logical structures are settled, the third Darwinian principle, selection, can be introduced as well. This role is played by the so called weak anthropic principle, or WAP [11], pointing out that only those universes can be observed where observers can appear, which selects a narrow class of fine-tuned universes as it is noted in Weinberg's quotation above. Thus, though our universe is thought of in this Darwinian approach as a random representative of the Tegmark's totality of forms, its fine tuning apparently receives a scientific explanation as a result of a Darwinian chaosogenesis. Although in the infinite megaverse only a tiny portion of universes is fine-tuned for life and consciousness, the probability for any observer to see the universe as fine-tuned is one hundred percent.

An important role of WAP as the only alternative to theistic explanations of the fine tuning was stressed by S. Weinberg [12]:

> In me, this apparent fine-tuning arouses wonder. The only explanation for it, other than a theological explanation, is in terms of a multiverse— I mean a universe consisting of many parts, each with different laws of nature and different values for its constants, like the 'cosmological constant' which governs cosmic expansion. If there is a multiverse consisting of many universes, most of them hostile to life but a few favorable to it, then it's not surprising that we find ourselves in one where conditions are in the fortunate range.

Nothing seemingly contradicts the assumption that our universe is a random representative of WAP-selected subset of Tegmark's multiverse, but is that really so? Does the universe indeed have no clear signature excluding any possibility of it having been randomly selected from this totality of all possible mathematical structures? Is the concept of chaosogenesis irrefutable by any thinkable observation, i. e. is it not a scientific hypothesis? Apparently, it is considered as irrefutable by some leading experts. For instance, Brian Greene clearly says that [13]:

> I draw the line at ideas that have no possibility of being confronted meaningfully by experiment or observation, not because of human frailty or technological hurdles, but because of the proposals' inherent nature. Of the multiverses we've considered, only the full-blown version of the Ultimate Multiverse falls into this netherland. If absolutely every possible universe is included, then no matter what we measure or observe, the Ultimate Multiverse [*i.e. Tegmark's one*] will nod and embrace our result.

Contrary to B. Greene, we are showing below that Tegmark's hypothesis runs counter to certain observations, so it fails, and fails as a scientific theory.

## Weak Anthropic Principle

On the question of possibility of the long evolution from the Big Bang to thought the WAP answers thus: in those worlds, where this path hasn't been traversed, there is no one to ask. Alright, in our universe this path has been traversed, so lets ask one more question: why isn't the path thrown into nowhere right now? Why does this world not only exist, but continues to exist, and the prediction of its continued existence comes true over and over, while the prediction of the end of the world turns out to be false again and again? What keeps this complex world with its life and thought in being?

The prediction of an immediate end of the world is completely unavoidable in the framework of the WAP and full-blown multiverse. Maintaining whatever special features is a special requirement, demanded of the universe. Special demands can be fulfilled, if appropriately grounded. If there is no ground, then there is no sense in expecting of keeping the requirements. The WAP explains why life and thought became possible. But out of the truism, which it uses to explain, no logical consequence follows that further on the required conditions will remain satisfied.

The reader might ask, if it already turned out this way with our universe, that up till now it maintained life, does it mean that it has some kind of a foundation of the laws that it happened to have, which keep it in this status of continuation of life. What's wrong with this explanation of the renewed anthropic continuation?

Let's consider this explanation more closely. It supposes an existence of some laws, giving structure to the universe, its evolution in time. The laws themselves at the same time must be atemporal: otherwise whatever segregation of them from the temporal world would be meaningless; the regulators would be no different than regulated. But even beside that they are atemporal, the laws are mental entities: to see them and to think them is one and the same. Postulating laws as objective mental entities implies Mind as a sphere of their being. It is this Mathematical Mind that differentiates the law good for the universe from one which is not – meaninglessness, absurdity or a self-contradictory system. Because the Mind not only discerns the laws from non-laws but it manifests them as structure-forming elements of the material universes, It is also a Maker. In this way, the very assumption of some non-contradictory laws leads to the conclusion about the transcendental Creator as the Mathematical Mind and Maker, even strictly within the framework of Tegmark's full-blown multiverse.

Let's assume that Tegmark's multiverse is just that ocean, a random drop of which is our universe; a chance limited by the WAP. Does this assumption mean that the conditions for life, satisfied for billions of years will continue to be satisfied in the next second? Does the belonging to the full-blown multiverse, strengthened by billions of years of good behavior serve as the ground to conclude that in the next second this behavior will continue to be good? There is no such ground here. If a mathematical function of a general form, a random representative of all possible functions, has been at zero so far, then we can only conclude that this specific quality will not continue to be maintained even in the very near future.

Tegmark's multiverse, determined by all non-contradictory sets of formulas, essentially is no different from a multiverse limited by nothing, that is pure chaos. Whatever the behavior of the universe up till now, there will always be an infinite number of laws corresponding to this behavior, and the chance to select out of them the set of laws that guarantees good behavior even for the next second equals to zero. There is an infinite number of laws of explosive action in Tegmark's multiverse, sleeping up to a certain moment and waking arbitrarily soon. To exclude the awakening of every one of this infinity of laws in the next second would equate to postulating an ungrounded specificity of our universe among the multiverse.

And so, just the conformity of the universe to smooth laws is not enough to conclude its good behavior in the nearest future. The unavoidable conclusion on this basis is the immediate end of the world. In order to avoid this conclusion, it is necessary to rule out laws of explosive action from the initial multiverse, because the truism of WAP does not exclude them.

What is left then of the original motivation of "not needing this hypothesis," of the Creator, which motivation is responsible for the WAP? The above analysis of implications demonstrates complete failure of that plan. According to those conclusions, to which we come unavoidably, the Ultimate Mind is necessary not only as a Mathematical one, guaranteeing non-contradiction of laws, but also as an Architectural one, limiting participation of laws of explosive action. It will be shown further that the laws of our universe point to yet one more substantial selection.

## A Cosmic Observer

"Observers" in WAP are not normally specified; it is not taken into account what it is namely they do observe. We suppose that to be qualified as "observers" they at least have to be conscious, as it is also reasonable to assume that conscious creatures observe their immediate space of life support and have access to at least empirical knowledge about it. However, this sort of knowledge has nothing to do with theoretical knowledge of the big cosmos; the first by no means entails the second. Let us fix this point of an important distinction, a distinction between those *simple, minimal, empirical observers* and *cosmic observers*, who are discovering theories of big cosmos, seeing their universe both at extremely large and extremely small scales, far exceeding the scale of immediate life support. To become cosmic observers, minimal ones must live in a very specific world among the populated worlds. Specifically, their universe has to be theoretically comprehensible on a big cosmic scale; their world has to be *theoretizable*, so to say. In other words, the possibility for observers to be not just simple but cosmic requires their universe to have a very special logical structure: it has to be described by elegant laws, covering many orders of magnitude of their parameters. Contemporary humanity is indeed a cosmic observer. For today, our scale of scientific cognition is described by an enormous dimensionless parameter $\sim 10^{45}$; that big is the ratio of the sizes of largest object of physics, the universe, $\sim 10^{26}$m, to the smallest ones, the top quark and the Higgs boson, corresponding to $\sim 10^{-19}$m.

## The Condition of Elegance

This condition of *theoretizability* apparently is extraneous to the selective anthropic principle, that is, theoretizability seems unnecessary for the universes to be populated by conscious creatures or to be observed. In fact, the latter condition is essentially local; it requires something like a life-friendly planet inside any universe. The former condition, though, is global; it requires the laws of nature to be elegant on a big cosmic scale, a scale by far exceeding that of the life on the planet. Generally, local conditions do not entail global consequences, and since theoretizability is a specific functional requirement detached from WAP selection, we have to conclude that it is highly unlikely for an observed universe to be theoretizable. Since we know, after Isaac Newton, that our universe is theoretizable, chaosogenesis theory is apparently refuted. However, this refutation, being qualitative only, leaves a possibility to object. Its core statement, that theoretizability is a specific requirement *detached* from the anthropic condition for universes "to be observed" can be questioned. How can we be sure that theoretizability is logically independent from WAP? It would not be independent, if WAP did not allow for our theoretizable laws of nature any visible modifications even at extremes of very large and very small scales, modifications that might exclude the appearance of conscious beings for one or another reason. The very concept of a fine-tuned universe is suggesting to us that sort of an idea concerning the fundamental constants, and so, we may ask: what if the same is true concerning the very structure of the laws of nature? Although it is hard to believe that moderate modification of, say, General Relativity at the distances exceeding the solar system, can dramatically reduce the possibility of consciousness on our planet, we should consider a chance that it cannot be excluded. This very argument for the strong relation between the weak anthropic principle and theoretizability was recently suggested by A. Linde [14]:

> ... the inflationary multiverse consists of myriads of 'universes' with all possible laws of physics and mathematics operating in each of them. We can only live in those universes where the laws of physics allow our existence, which requires making reliable predictions.

The same idea was expressed in the latest book of M. Tegmark [15], with reference to E. Wigner:

> An anthropic-selection effect may be at work as well: as pointed out by Wigner himself, the existence of observers able to spot regularities in the world around them probably requires symmetries, so given that we're observers, we should expect to find ourselves in a highly symmetric mathematical structure. For example, imagine trying to make sense of a world where experiments were never repeatable because their outcome depended on exactly where and when you performed them. If dropped rocks sometimes fell down, sometimes fell up and sometimes fell sideways, and everything else around us similarly behaved in a seemingly random way, without any discernible patterns or regularities, then there might have been no point in evolving a brain.

Let's accept this arguable hypothesis, and suppose that somehow WAP does not allow significant deviations of the laws of nature locally compatible with conscious beings from the globally theoretizable form. Then, the question is: which deviations from the existing laws are allowed by the anthropic principle? If the world is generated by chaos, all imaginable additional terms to the life-selected ones are coming into play; the amplitude or width of the resulted deviation is limited by the anthropic principle, but functional behavior of the deviation is arbitrary. We have some estimations about the allowed deviation in the context of fine-tuning as relative variations of the fundamental constants compatible with WAP, and the most stringent of them are at the order of $10^{-3}$, i.e. 0.1% [11]. Since we are considering here the problem of functional accuracy, the enormously stringent requirements on some constants, like the initial conditions at the big bang [16], do not reduce the amplitude of these functional variations. Thus, working on the Linde argument, we may roughly estimate the sensitivity of the anthropic selection to the relative functional variations of the fundamental laws to not be finer than 0.1% or so. If the laws of nature were generated by a random choice from Tegmark's multiverse, they would be expressed by irregular functions possibly following elegant ones within a relative width of ~0.1% or more. In this respect, it does not matter whether chaos reveals itself through arbitrary functions or arbitrary mathematical structures; with Tegmark's "mathematical democracy" functional representatives of the two families are indistinguishable and are dominated by extremely complicated, practically irregular functions. The elegant formulas might be approximations to the real irregular fundamental laws with that WAP-determined accuracy, but not better.

Moreover, measurements of the fundamental constants in this world would be reproducible only at the anthropic level, not better. If physicists of that hypothetical world tried making measurements of their fundamental constants at the better accuracy, they would realize that none of the measurements are reproducible at that level; they would all contain space-time noise with a relative amplitude of 0.001, driven by infinitely complicated terms of the true laws of nature. So, physics in that Tegmarkian universe would be stopped at the anthropic accuracy level simply because, with the probability of 100%, no reproducible measurement would be possible there with accuracy better than that.

We know though, that the real accuracy of our fundamental theories is not only better than anthropic, but many orders of magnitude better; they are absolutely precise on that scale. Indeed, the General Relativity test with a double neutron star PSR 1913+16 showed an unprecedented agreement between theory and observation at the level of $10^{-14}$. Another impressive demonstration of that extremely high precision relates to the Quantum Electrodynamics: the theoretically predicted value of an electron's magnetic moment is confirmed by measurements with the accuracy ~$10^{-11}$; see e. g. Ref. [16]. Thus, many experiments which proved high precision of our elegant laws of nature, orders of magnitude better than the anthropic width, refute Tegmark's hypothesis.

This consideration shows, by the way, that cosmological chaosogenesis is a scientific hypothesis since it is falsified by observations. Note that the idea of the multiverse was at least partly motivated by the wish to find a pure scientific explanation to the fact of the fine-tuned universe: if our universe is the only one, its fine-tuning does not suggest any other reasonable explanation but an act of purposeful creation. For a single universe, its fine-tuning is too stringent for a purely scientific explanation, but the idea of multiverse chaosogenesis, suggested as an attempt to explain fine tuning within bounds of science, is refuted by the opposite reason: the estimated anthropic limitations on fine-tuning aren't anywhere fine enough to explain the experimental confirmations of the extreme precision of the elegant forms as fundamental laws.

## A Pythagorean Universe

After having announced the "complete mathematical democracy" at the beginning of his article, later on Tegmark notices that "our physical laws appear relatively simple". At this point, to be consistent with reality, he gives up the proclaimed "mathematical democracy" in favor of an aristocracy of simple mathematical forms. After such an overturn, his multiverse now has almost nothing to do with chaos; instead, it is generated by some source of elegant mathematical forms. As a result, "the embarrassing question" about the source of this ontological inequality of the mathematical forms remains as it was. This contradiction of Tegmark's "democratic" intention with his "aristocratic" practice was noted by Alex Vilenkin [17]:

Tegmark's proposal, however, faces a formidable problem. The number of mathematical structures increases with increasing complexity, suggesting that "typical" structures should be horrendously large and cumbersome. This seems to be in conflict with the simplicity and beauty of the theories describing our world.

Since the laws of our universe are not picked randomly, they can only be purposefully chosen. Our universe is special not only because it is populated by living and conscious beings but also because it is theoretizable by means of elegant mathematical forms, both rather simple in presentation and extremely rich in consequences. To allow life and consciousness, the mathematical structure of laws has to be complex enough so as to be able to generate rich families of material structures. From the other side, the laws have to be simple enough to be discoverable by the appearing conscious beings. To satisfy both conditions, the laws must be just right. The laws of nature are fine-tuned not only with respect to the anthropic principle but to be discoverable as well. Multiple aspects of this double fine tuning are discussed in Ref. [18] and references therein. In other words, the Universe is fine–tuned with respect to what can be called as the *Cosmic Anthropic Principle*: its laws are purposefully chosen for the universe to be *cosmically observed*. It could be even that our laws are at their simplest within our sort of life. Would it be possible to take any part away from our existing theories without compromising forms of life as we know them? Such a special universe deserves a proper term, and we do not see a better choice than to call it *Cosmos or* to qualify it as *Pythagorean*, in honor of the first prophet of theoretical cognition, who coined such important words as *cosmos* (order), *philosophy* (love of wisdom), and *theory* (contemplation).

Since chaosogenesis, being limited only by the anthropic principle, is the only option for a completely scientific solution of the problem of cosmogenesis, its refutation entails that the problem of cosmogenesis cannot be solved within the framework of science. Any scientific approach to that would require a specific set of axioms, consistent not only with the anthropic principle but with elegant mathematical forms truly underlying our world; however, we've shown that the question about embedment of one instead of another specific set as a logical structure of the universe cannot be scientifically answered.

Starting with Pythagoras, it was a matter of faith for sparse groups of few people and lonely individuals that "fundamental laws of nature are described by beautiful equations." Theoretical science was conceived and nurtured by this very faith with its "cosmic religious feeling", which inspired scientific cognition for twenty-five centuries. Without any exaggeration, all great theories, from those of Copernicus, Kepler and Newton to those of Einstein and Dirac happened as guesses on the grounds of some fundamentally simple ideas like symmetry, conservation, or equivalence. Likewise, Wigner saw "the appropriateness of the language of mathematics for the formulation of the laws of physics" as a miracle and "a wonderful gift which we neither understand nor deserve" [19]. His maxim "we should be grateful for it" can only have meaning if a mind to be grateful to is implied.

The noted forty-five orders of magnitude of scientific cognition, with more than ten digits of precision reached in some experimental verifications, allow us to conclude about a scientific confirmation of what was considered a matter of faith for two and a half millennia: now it is a matter of fact that the universe is indeed Pythagorean. In other words, the existence of the Platonic world of elegant mathematical forms structuring the physical world is scientifically confirmed, and the accuracy of this confirmation is many orders of magnitude better than that of any specific statement of physics.

After two and a half millennia since its birth, fundamental science reached a grade of maturity allowing for a dual confirmation of its faith: the Pythagorean faith is confirmed as prophecy coming true and as a good tree that brings forth good fruit.

## Three Worlds, Two Totalities

Pythagorean forms of the discovered laws of nature tell us that the ultimate goal of fundamental physics,

the theory of everything, either contains a significant Pythagorean core, or, what is more reasonable to assume, is totally Pythagorean. This Pythagorean core has to be powerful enough to generate a sufficiently rich set of Pythagorean laws, as we observe, but whatever this theory of everything is, it cannot be the ultimate answer to the question about the order of being, because this form is special due to there being other forms, and so, like any other, it does not constitute a totality. For laws of nature, there are only two thinkable explanatory principles, opposites of each other, which are totalities: chaos and mind as such.

Because the logical structure of our universe can not be explained by chaos, and because it can not explain itself, we are left with only one possible explanation remaining, that it was conceived and realized by a mind. A. Vilenkin prefers to formulate this apparently inevitable conclusion about the cosmic Mind as a question [17]:

> … the laws should be "there" even prior to the universe itself. Does this mean that the laws are not mere descriptions of reality and can have an independent existence of their own? In the absence of space, time, and matter, what tablets could they be written upon? The laws are expressed in the form of mathematical equations. If the medium of mathematics is the mind, does this mean that mind should predate the universe?

To be a complete terminus of questioning, a creative mind has to be mind per se, or the Absolute Mind. Otherwise, questions about origin and possibility of its mindness would require new answers. Unlike chaos, Absolute Mind as terminus leaves room for mystery; the creativity of the human mind does as well. Where there is mystery, questioning is inexhaustible, and the feeling of mystery instills a deep value in the pursuit of knowledge. Contrary to this, the postulation of chaosogenesis, by rejecting the primacy of mind, is incompatible with mystery, and thus with the value of fundamental cognition. Thus, the problem of cosmogenesis leads to a dual mystery, one aspect of which is the Absolute Mind as the source of the laws of nature, while the other aspect lies in a mind capable of discovering them. From this point of view, Tegmark's multiverse obtains a new meaning; it is a space for the search for interesting worlds to be created, with laws open to discovery.

It seems important to mention here that chaos, refuted as a possible source of the laws of nature, can and does participate in the physical world as indeterminism, by means of uncertainty left by the quantum laws of nature.

The very idea of observation, being so far associated with material objects only, is enriched by an even more fundamental meaning of the Platonic observation, i. e. observation of elements of the Platonic world structuring the material world. Cosmic observation is possible only due to a combined vision of both worlds. Roger Penrose suggested the idea and the image of "Three Worlds, Three Mysteries" [16]. The three worlds, Physical, Platonic, and Mental, differ time-wise. The Platonic world does not have any age at all; it is atemporal. The Physical world is temporal, and its age, counted from the border of all observations, the big bang, is calculated at $13.798 \pm 0.037$ billion years (note the precision!). The age of humanity as cosmic observers is extremely short on that scale. Yet although the history of many scientific discoveries is known minutely, and although we cannot observe anything closer than our thoughts, the genesis of the cosmic observer remains no less mysterious to us than the genesis of Physical and Platonic worlds.

Wonder of Pythagorean harmony of the fundamental laws of nature and continuing demonstrations of human ability to discover them so remotely from our own natural scale leads now more than ever before to deep questions about the three mysteries, whose entanglement and coherence are revealing the underlying Unity, the ultimate transcendental Source of everything existent, including ourselves, the growing cosmic observers.

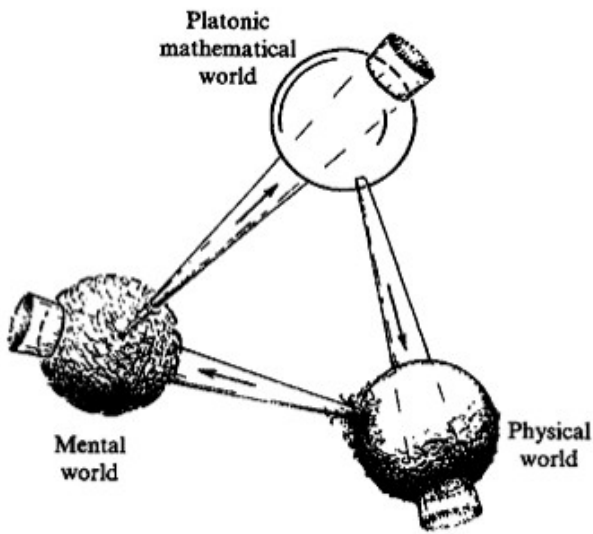

R. Penrose, "Three Worlds, Three Mysteries" [16].

[1] L. Wittgenstein, "Tractatus Logico-Philosophicus" (1921)

[2] P. Davies, "The Goldilocks Enigma", Houghton Mifflin Harcourt. Kindle Edition (2006).

[3] P. Davies "How bio-friendly is the universe". *Int.J.Astrobiol* **2**, p.115 (2003).

[4] S. Hawking, "A Brief History of Time", Bantam Books, p. 125 (1988)

[5] A. Tsvelik, "Life in the Impossible World", Ivan Limbakh publ., St.-Petersburg, 2012 (in Russian).

[6] K. Jaspers, "The Great Philosophers", A Harvest Book, Vol.2 (1966).

[7] M. Tegmark, "The Mathematical Universe", Foundations of Physics 38 (2), p. 101 (2007).

[8] S. Weinberg, "Lake Views: This World and the Universe" (Kindle Locations 337-338). Harvard University Press. Kindle Edition (2011).

[9] L. Smolin, "The Life of the Cosmos" (1997)

[10] A. Linde, "The Self-Reproducing Inflationary Universe", Sci. Am., Vol. 271, No. 5, pp 48-55, (1994)

[11] J.D. Barrow, F.J. Tipler "The Anthropic Cosmological Principle." (1988)

[12] J. Holt, "Why Does the World Exist?: An Existential Detective Story" (pp. 156-157). Liveright. Kindle Edition.

[13] B. Greene, "The Hidden Reality: Parallel Universes and the Deep Laws of the Cosmos" (Kindle Locations 5943-5947). Knopf Doubleday Publishing Group. Kindle Edition (2011).

[14] A. Linde "Why Is Our World Comprehensible?", in "This Explains Everything", Ed. J. Brockman, 2012.

[15] M. Tegmark, "Our Mathematical Universe: My Quest for the Ultimate Nature of Reality", Knopf Doubleday Publishing Group. Kindle Edition.

[16] R. Penrose, "The Road to Reality", First Vintage Books Edition, 2007.

[17] A. Vilenkin, "Many Worlds in One: The Search for Other Universes" (Kindle Locations 3188-3191). Farrar, Straus and Giroux. Kindle Edition (2006).

[18] G. Gonzalez and J. W. Richards, "The Privileged Planet", Regnery Publishing, Inc (2004).

[19] E. Wigner, "The Unreasonable Effectiveness of Mathematics in the Natural Sciences", Communications on Pure and Applied Mathematics, vol. 13, issue 1, pp. 1–14 (1960).